\documentclass[%
 reprint,
 amsmath,amssymb,
 aps,
floatfix
]{revtex4-1}

\usepackage{graphicx}% Include figure files
\usepackage{dcolumn}% Align table columns on decimal point
\usepackage{bm}% bold math
\usepackage{xcolor}
\begin{document}

%\preprint{APS/123-QED}

\title{Terahertz third harmonic generation in c-axis La$_{1.85}$Sr$_{0.15}$CuO$_4$}

\author{Kelson Kaj$^1$}
\thanks{These two authors contributed equally}
\author{Kevin A. Cremin$^1$}
\thanks{These two authors contributed equally}
\author{Ian Hammock$^1$}
\author{Jacob Schalch$^1$}
\author{D. N. Basov$^2$}
\author{R. D. Averitt$^1$}
\email{Corresponding author email: raveritt@ucsd.edu}
\affiliation{
$^1$University of California, San Diego, CA \\
$^2$Department of Physics, Columbia University, New York, NY
}

%\date{\today}

\begin{abstract}
Terahertz nonlinear optics is a viable method to interrogate collective phenomena in quantum materials spanning ferroelectrics, charge-density waves, and superconductivity. In superconductors this includes the Higgs amplitude and Josephson phase modes. We have investigated the nonlinear c-axis response of optimally doped La$_{1.85}$Sr$_{0.15}$CuO$_4$ using high-field THz time domain spectroscopy (THz-TDS) at field strengths up to $\sim$80 kV/cm. With increasing field, we observe a distinct red-shift of the Josephson plasma edge and enhanced reflectivity (above the plasma edge) arising from third harmonic generation. The non-monotonic temperature dependent response is consistent with nonlinear drive of the Josephson Plasma Mode (JPM) as verified with comparison to theoretical expectations. Our results add to the understanding that, using THz light, the JPM (in addition to the Higgs mode) provides a route to interrogate and control superconducting properties. 

\end{abstract}

\pacs{Valid PACS appear here}% PACS, the Physics and Astronomy
\maketitle

%\tableofcontents

\section{Introduction}
\label{sec:Intro}

 Terahertz spectroscopy has emerged as a powerful probe of non-equilibrium dynamics in quantum materials \cite{Zhang2014,delaTorre2021}. More recently, the generation of intense terahertz (THz) high-field pulses has enabled nonlinear drive of the low energy electrodynamics, offering new insights into the many-body response while also providing a route for on-demand control of emergent properties \cite{Kampfrath2013,Liu2012,Nova2019,Basov2017}. Superconductors are particularly amenable to high-field interrogation and manipulation of the condensate with terahertz pulses. This includes the Higgs amplitude mode in conventional superconductors where light couples non-linearly to the condensate, driving oscillations of the order parameter amplitude at twice the pump frequency, leading to harmonic generation \cite{Matsunaga2014,Matsunaga2017,Cea2016,Pekker2015,Tsuji2015,Katsumi2018,Shimano2020}. Higgs mode spectroscopy has been extended to cuprate superconductors with novel mode dynamics associated with the d-wave gap symmetry and to iron pnictides where multiband effects have been observed \cite{Chu2020,Vaswani2021}. 
 
 %Then the next paragraphs can be Josephson as an additonal access route.

\begin{figure}
    \centering
   \includegraphics[width=7cm]{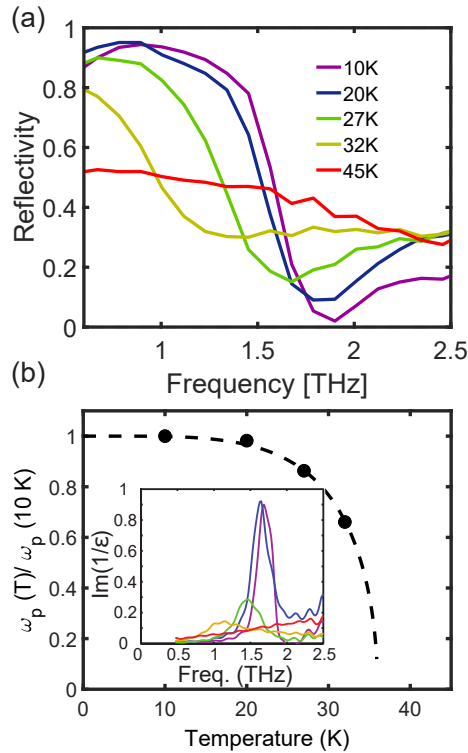}
    \caption{(a) Low field c-axis reflectivity of  La$_{1.85}$Sr$_{0.15}$CuO$_4$ at several temperatures. At 10K, $\omega_p$ is $\sim$1.7 THz.  (b)  Temperature dependence of $\omega_p$ below $T_c$ normalized to the low temperature measurement $\omega_p (T= 10$ K). The dashed line is a fit to a BCS-like order parameter temperature dependence, as described further in the text.   The inset shows the (normalized) loss function, -Im$(1/\epsilon)$, using the same legend in panel (a).}
    \label{fig:LSCO_lowfield_ref}
\end{figure}

In the high-$T_c$ cuprates the phase mode response (Josephson plasma mode -- JPM) manifests at THz frequencies \cite{Tamasaku1992,Basov2005}. Briefly, the copper-oxygen planes are weakly interacting and Josephson coupling dictates c-axis Cooper pair tunneling based on the interlayer phase difference of the superconducting order parameter between adjacent planes. This results in a plasma edge in the c-axis reflectivity at the Josephson plasma frequency $\omega_{p}$. For single layer cuprates, $\omega_{p}$ typically manifests at THz frequencies with $\omega_{p}^{2}$ proportional to the condensate density $n_{s}$ (e.g., see Fig.\ref{fig:LSCO_lowfield_ref}, discussed in more detail below). As such, the JPM serves as a reporter of the condensate response which includes nonlinearities such as field-induced renormalization of $\omega_{p}$ and harmonic generation \cite{Savelev2010,Rajasekaran2016,Rajasekaran2018,Gabriele2021}. 

%and exhibits abundant nonlinearities that are of intrinsic interest and unveil phase competition in cuprates, with La$_{2-x}$Ba$_{x}$CuO$_{4}$ as a prime example.

We investigate the nonlinear spectral response of c-axis La$_{1.85}$Sr$_{0.15}$CuO$_4$ (LSCO) using high field THz-TDS as a function of field strength and temperature.  
A redshift of $\omega_{p}$ with increasing field (2.4 kV/cm up to 80 kV/cm) arises from the Josephson effect. With increasing temperature the maximum redshift increases from $\sim$110 GHz at 10 K $\sim$220 GHz at 32 K. This temperature dependent frequency shift cannot be explained solely using the Josephson equations which predict a high-field shift of the JPM is a constant fraction of the equilibrium JPM frequency at each temperature. This is the opposite of the observed behavior, and could be related to increased quasiparticle damping at higher temperatures. Commensurate with this is broadband third harmonic generation above the plasma edge which exhibits a slight increase with decreasing temperature below T$_{c}$, dropping off in the normal state. The temperature dependence is compared with calculations based on the theory in Reference \cite{Gabriele2021}. The qualitative agreement between calculations and experiment suggests that the temperature dependence is related to the competing factors of Josephson coupling strength, the resonance of the pump with the JPM, thermal population of excited plasmon states, and quasiparticle damping. For our broadband drive, we estimate a power conversion efficiency of $\sim$ 6$\times$10$^{-5}$.

\section{Methods}
\label{sec:Methods}

THz radiation is generated via optical rectification using a Ti:sapphire regenerative amplifier (1 KHz, 800 nm, 100 fs, 3 mJ) using tilted pulse front generation in a Mg-LiNbO$_3$(LNO) crystal  \cite{Hebling2002,Hebling2008}.  The THz output from the LNO crystal surface is collimated with a lens ($f$ = 120 mm) and focused onto the sample with an angle of incidence of 15$^{\mathrm{o}}$ and a beam diameter of $\sim$2.3 mm FWHM (pulse energy $\sim$2 $\mu$J). Before reaching the sample, the THz light passes through a pair of wire grid polarizers which are used to attenuate the THz pulse, covering the range from 2 - 80 kV/cm. The reflected beam from the sample surface is collected and focused onto a 300 $\mu$m thick (110) GaP crystal for EO sampling. The GaP crystal is mounted on a 2 mm thick (100) GaP crystal to delay the internal etalon inside the GaP to times beyond our temporal measurement window. Roughly 1\% of the pump beam is split off and used for gating the THz pulse in the GaP crystal. The entire THz beam path is in a vacuum chamber (schematic of the experimental setup is shown in Fig. 2a). The bandwidth of the pulses extends from 0.2 - 3 THz (see Fig. 3a) with a maximum at $\sim$0.65 THz. The full spectral content is used for the nonlinear studies and is not spectrally filtered.

The LSCO crystal was cut and polished to expose the a-c plane and was grown via a traveling-solvent floating-zone method \cite{Tanaka1989} with a surface size of approximately 3mm $\times$ 3mm. Reflectivity measurements were performed by taking time domain scans of the electric field reflected off the sample and from a gold reference mirror at the sample position in both nonlinear and linear regimes. The reflectivity was then obtained by calculating the Fourier transforms of the time domain scans and taking their ratios, $R_{NL} = | E_{NL}(\omega)/  E^{Au}_{NL}(\omega)|^2$. Measurements were performed above and below $T_c$ = 38K as described below.

\section{Results}
\label{sec:Results}

We first measured the temperature dependent c-axis response at the lowest available electric field to characterize the linear response. The c-axis reflectivity is plotted in Fig. \ref{fig:LSCO_lowfield_ref}(a) for several temperatures above and below $T_c = 38$ K, and shows a clear plasma edge emerge and sharpen with decreasing temperature from 32 K to 10 K, coinciding with an increase in superconducting condensate. The low temperature (10K) plasma edge $\omega_p$ is at $\sim$1.7 THz is in agreement with previous c-axis measurements for $x$ = 0.15 doping \cite{Tamasaku1992,Dordevic2003}. Figure \ref{fig:LSCO_lowfield_ref}(b) shows the plasma frequency $\omega_p (T)$ normalized by the low temperature measurement $\omega_p$(10 K) and scales with BCS-like order parameter temperature dependence, as $\omega_p^2\propto \mathrm{tanh}(2.26\sqrt{T_c/T-1}))$ \cite{Sim2017,Yuan2019}. The inset of Fig.\ref{fig:LSCO_lowfield_ref}(b) displays the normalized loss function -Im($1/\epsilon)$ at each temperature. 

\begin{figure*}
    \centering
    \includegraphics[width=0.9\textwidth]{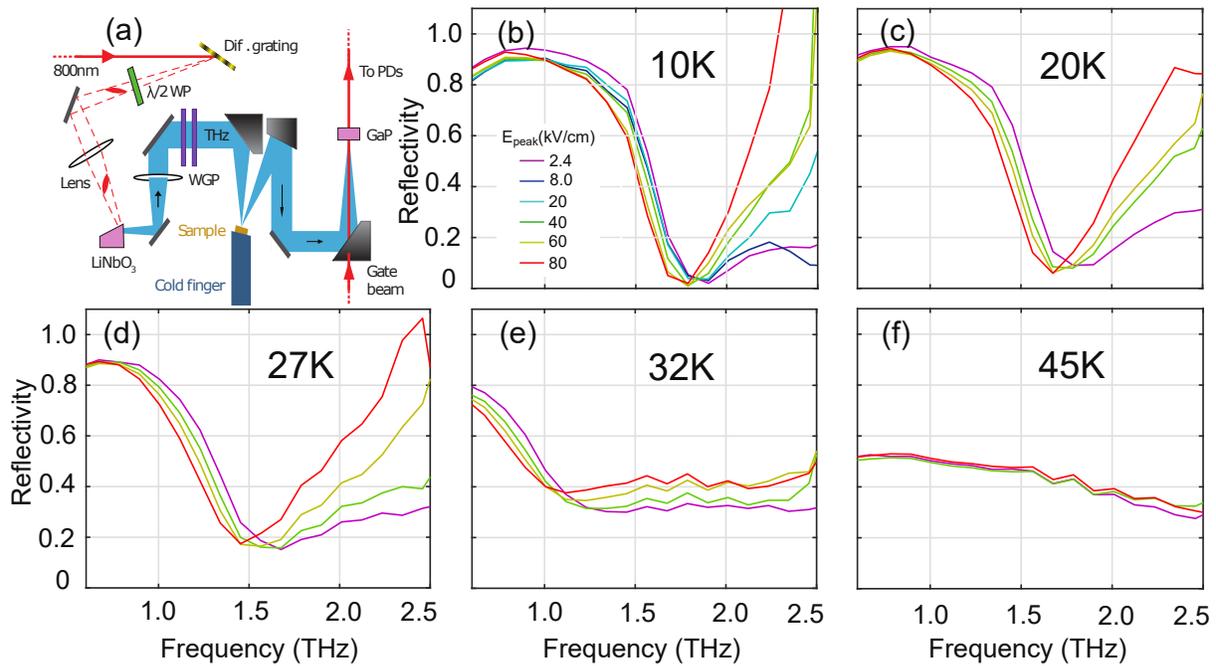}
    \caption{La$_{1.85}$Sr$_{0.15}$CuO$_4$ c-axis THz reflectivity. (a) Schematic of experimental setup. Panels (b)-(f) show the reflectivity taken at temperatures 10 K, 20 K, 27 K, 32 K, and 45 K respectively. At each temperature, the reflectivity was measured for fields ranging from 2.4 -- 80 kV/cm as indicated in the legend.}
    \label{fig:LSCO_HighField_allTemps}
\end{figure*}

For frequencies below $\omega_p$ the reflectivity is $\sim$90\%, whereas near-unity reflection is expected as $\omega \rightarrow 0$. This deviation is attributed to imperfect referencing since the sample size and beam diameter are comparable. This effect is more pronounced at lower frequencies as the focused THz beam diameter is frequency dependent. However, the low temperature $\omega_p (T)$ (and associated temperature dependence) is consistent with previous studies for $x$ = 0.15 doping, indicative of a high-quality crystal. The reflectivity presented in Fig. 1 and 2 are normalized to match reflectivity measurements using Fourier transform infrared spectroscopy (FTIR) on the same LSCO crystal at 1 THz \cite{Schalch2019}.

The nonlinear terahertz reflectivity results are shown in Fig. 2b-f. As shown in Fig. 2b (base temperature 10 K), two pronounced effects occur: There is a redshift of the plasma edge with increasing field and, above the plasma edge, there is an increase in the reflectivity corresponding to third harmonic generation (as described in greater detail below). With increasing temperature, there is a decrease in the condensate density, but there is still a clear redshift in the plasma edge and enhanced reflectivity. At 32 K (Fig. 2e), the plasma edge shift as a function of field is $\sim$220GHz, larger than the $\sim$110GHz shift at 10 K, but the reflectivity increase is relatively small. Above T$_{C}$ (Fig. 2f), there is no longer a superconducting response and the nonlinear increase in reflectivity is minimal to nonexistent. The nonlinear reflectivity data in Figure 2 is informative as it reveals both the JPM redshift and the increase in reflectivity above $\omega_p$. However, the increase in reflectivity requires a more careful analysis as we now discuss.

Figure 3a plots the spectral amplitude of the electric fields reflected from the LSCO at 10K, comparing the reflectivity of the nonlinear (NL) and linear (L) response. The red curve is E$_{NL}(\omega)$ at the highest field (80kV/cm). The black curve is E$^{*}_{L}(\omega)$ at the lowest field, appropriately normalized to enable quantitative comparison with the spectral changes that arise in the nonlinear regime \cite{Scale_note}.

As with the reflectivity data in Figure 2, the redshift of the plasma edge is evident when comparing  E$^{*}_{L}(\omega)$ and E$_{NL}(\omega)$ in Fig. 3a. Moreover, the increase in the spectral amplitude above the plasma edge is also evident (blue shaded region). The peak spectral amplitude of the terahertz pulses is at $\sim$0.65 THz. The second peak in E$_{NL}(\omega)$ is at $\sim$1.9 THz, consistent with third harmonic generation. The data in Fig. 3a reveals that the nonlinear signal is relatively small in comparison to the peak amplitude of the incident pulse (i.e.,E$^{*}_{L}(\omega)$ at 0.65 THz), but considerably larger than the spectral spectral amplitude of E$^{*}_{L}(\omega)$ between 2 - 3 THz. Nonetheless, the dynamic range is sufficient to enable a determination of the field dependence of the integrated spectral amplitude above the plasma edge (Fig. 3b), which further verifies that the enhanced signal arises from third harmonic generation (THG).

\begin{figure}
    \centering
    \includegraphics[width=6cm]{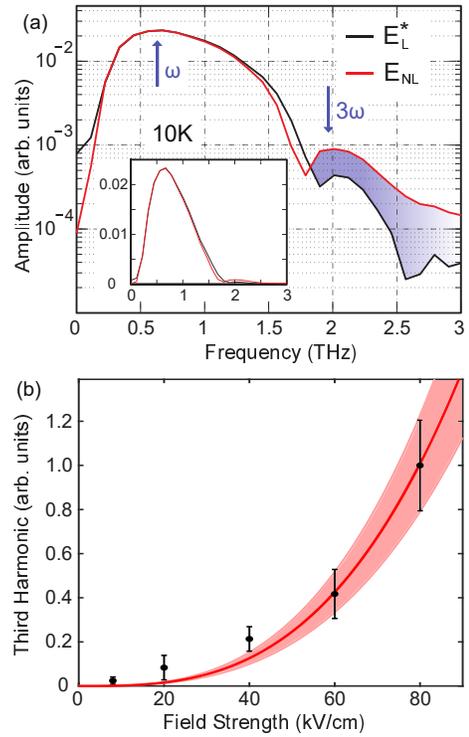}
    \caption{Third harmonic generation from $c$-axis LSCO at 10K. (a) Linear (black) and nonlinear (red) spectral amplitude of the THz pulses reflected from the sample. The nonlinear spectrum is at the maximum field strength of $\sim$80 kV/cm, and the linear spectrum is at the minimum field strength (renormalized as explained in the text). The shaded region in blue is the spectral content attributed to third harmonic generation. The inset contains a plot of the spectral amplitudes with a linear y-axis. (b) Third harmonic generation as a function of incident THz field strength at 10K. Each data point is from integrating Eq. 1 over the third harmonic region. The red line is a cubic polynomial fit, and the shaded area is bounded by cubic fits of the data including the upper and lower bounds of the data including the error bars.}
    \label{fig:LSCO_THG_Scaling}
\end{figure}

The THG response in Fig. 3(b) is quantified by integrating
\begin{equation}
    \label{eq:THG_integral}    
    E_{3\omega} \equiv \int \bigg[ E_{NL}(\omega) - E^{*}_{L}(\omega)\bigg] d\omega
\end{equation}
where E$_{NL}(\omega)$ and E$^{*}_{L}(\omega)$ are the reflected signals from the sample as defined above. To avoid integrating over the Josephson plasma edge, the integration is taken from 1.8 THz to 2.4 THz. Figure \ref{fig:LSCO_THG_Scaling}(b) displays the magnitude of the third harmonic $E_{3\omega}$ as a function of field strength at 10 K, along with a cubic polynomial fit (red line, $R$-squared value of 0.99). This further confirms the reflectivity/spectral amplitude increase arises from third harmonic generation in the superconducting phase. The lowest field strengths have more third harmonic intensity than the cubic fit, which may be attributed to the fit over-favoring the highest fields.

\section{Discussion}

The experimental observations can be further understood by examining the higher order terms in the phase dynamics of layered superconductors described by the Josephson equations \cite{Savelev2010,Laplace2016a}. 

In a layered superconductor, the interlayer phase difference $\theta (t)$ changes with time according to the second Josephson equation

\begin{equation}
    \label{eq:Josephson_II}
    \frac{\partial \theta(t)}{\partial t} = \frac{2edE(t)}{\hbar }
\end{equation}
where $2e$ is the cooper pair charge, $d$ is the interlayer spacing ($\sim$1 nm), $\hbar$ is Planck's constant divided by $2\pi$, and $E(t)=E_0 \mathrm{sin}(\omega_{\mathrm{pump}}t)$ is an electric field along the c-axis with $E_0$ the field strength which oscillates at frequency $\omega_{\mathrm{pump}}$. Solving equation \ref{eq:Josephson_II} yields the relation $\theta(t) = (2edE_0/\hbar)\mathrm{cos}(\omega_{\mathrm{pump}}t)$. Since the c-axis superfluid density $\rho_c$ scales as the order parameter phase difference, $\rho_c \propto \mathrm{cos}\theta$ and $\rho_c \propto \omega_p^2$ as shown in Fig. \ref{fig:LSCO_lowfield_ref}(b), the plasma frequency renormalizes according to $\omega_{NL}^2=\omega_p^2 \mathrm{cos}\theta(t)$ where $\omega_{NL}$ is the new Josephson plasma frequency under intense field strengths. By inserting $\theta(t)$ and expanding the $\omega_{NL}$ we arrive at 

\begin{equation}
\label{eq:JPR_redshift}
\begin{split}
    \omega_{JPM}^2 & = \omega_{JPM0}^2 \mathrm{cos}(\theta) = \omega_{JPM0}^2\mathrm{cos}\bigg[\theta_0 \mathrm{cos}(\omega_{pump}t)\bigg] \\
    & \approx \omega_{JPM0}^2 \bigg[ 1 - \frac{\theta_0^2}{4} - \frac{\theta_0^2 \mathrm{cos}(2\omega_{pump}t)}{4} + \dots \bigg] 
\end{split}
\end{equation}
where $\theta_0 = 2edE_0/\hbar$. From this expansion we can see that the next leading order term is subtracting, resulting in a redshift in the plasma frequency. This is what is observed below $T_c$ for high fields as shown in Fig. \ref{fig:LSCO_HighField_allTemps}.

The tunneling interlayer current depends on the phase difference as $I(t)=I_0\mathrm{sin}[\theta(t)]$, and solving for $I(t)$ gives

\begin{equation}
\label{eq:THG_current}
\begin{split}
    I(t) & = I_c \mathrm{sin}\bigg[ \theta_0 \mathrm{cos} (\omega_{\mathrm{pump}}t) \bigg] \\
         & \approx I_c \bigg[\theta_0 \mathrm{cos}(\omega_{\mathrm{pump}}t) - \frac{\theta_0^3}{6} \mathrm{cos}^3(\omega_{\mathrm{pump}}t) + \dots \bigg]
\end{split}
\end{equation}
where the leading higher order in the expansion is cubic. This expanded term leads to driving the current at the third harmonic and is observed as THz emission at $3\omega$, which manifests as a reflectivity increase above the plasma edge as shown in Fig. \ref{fig:LSCO_HighField_allTemps}(a)-(d).

\begin{figure}
    \centering
    \includegraphics[width=7cm]{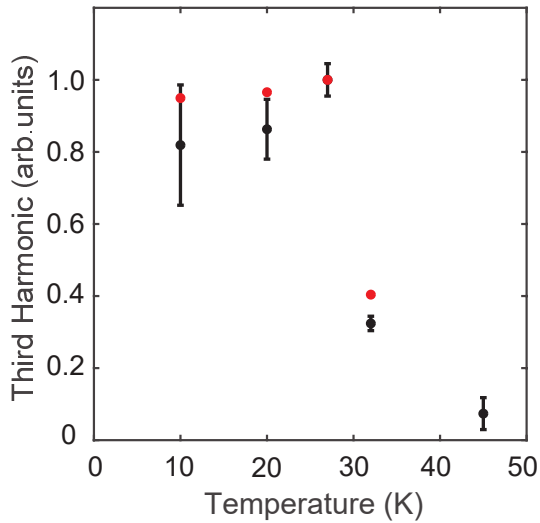}
    \caption{Magnitude of c-axis THG from LSCO versus temperature. All data points were taken with maximum field strengths of 80 kV/cm. The black circles are experimental data and the red circles are calculations as described in the text. The experimental data and calculations are normalized to the maximum value, which occurs at 27K in both experiment and simulation.}
    \label{fig:TH_100P_vs_Temp}
\end{figure}

The above equations predict the third harmonic signal to scale with the superfluid density, as observed in previous work on LBCO \cite{Rajasekaran2018}. The temperature dependence of the third harmonic emission for our LSCO studies is shown in Fig. \ref{fig:TH_100P_vs_Temp} (black dots). Clearly, the signal does not solely scale with the superfluid density (which is directly proportional to the square of the JPM frequency), but instead decreases slightly for temperatures lower than 27K. Recent work has shown that detailed calculations are necessary to understand the temperature dependence of the third harmonic signal. Calculations based on this theory are plotted as red dots in Fig. 4\cite{benf2021}. The third harmonic signal is calculated from the nonlinear optical kernel of the Josephson phase, and its overlap with the spectral amplitude of the pump pulse, as shown in equation \ref{eq:kernel}.  
\begin{equation}
\label{eq:kernel}
\begin{split}
    I^{NL}(\omega)=\int d(\omega') A(\omega-\omega') K(\omega') A^2(\omega')\\
     K\propto \frac{\omega_J^3 coth(\beta \omega_J/2)}{4\omega_J^2-(\omega+i \gamma(T))^2}
\end{split}
\end{equation}

\noindent $A$ is the spectral amplitude of the electric field profile and $K$ is the nonlinear optical Kernel. This kernel does have an overall scaling with the superfluid density as $\omega_J^3$.  However, there is also a factor in the kernel, $coth(\beta \omega_J/2)$, that comes from the thermal excitation of plasmon modes, which causes third harmonic emission to increase with temperature \cite{Gabriele2021}. The kernel also has a resonance at the Josephson plasma frequency. For our experiment, the pump pulse is centered at a frequency lower than the Josephson plasma frequency for all measured temperatures. However, with increasing temperature this process is closer to being on resonance since the condensate density (and hence $\omega_{p}$) is reduced. Finally, quasiparticle damping increases with temperature, causing third harmonic emission to decrease. Thus, there are four competing factors that determine the overall temperature dependence of third harmonic emission due to Josephson plasma waves. These calculations capture the qualitative features of the data including a maximum in third harmonic emission at 27K, a decrease as the temperature is decreased, and a sharp decline in third harmonic emission in the vicinity of T$_{C}$.

The work in \cite{Gabriele2021} and the results shown here motivate looking at the temperature dependence of third harmonic emission from other c-axis cuprates since the non-monotonic temperature dependence goes beyond basic Josephson-equations predictions. To our knowledge, third harmonic emission from c-axis cuprates has only been previously reported in La$_{2-x}$Ba$_{x}$CuO$_{4}$\cite{Rajasekaran2018} and shares some similarities with our results. More detailed studies of third harmonic generation and THz nonlinearities have been performed in many superconductors, including cuprates, focusing on light polarized in the ab-plane. The interpretation of the data has been in terms of the Higgs mode and quasiparticle contributions \cite{Chu2020,Chu2021,Matsunaga2012,Nakamura2018,Matsunaga2014,Matsunaga2017}. We note that there is a prediction of contribution of plasma waves to the third harmonic generation for light polarized in the ab-plane \cite{Gabriele2021}. The nonlinear THz response of cuprates can also be studied with pump-probe protocols, with \cite{Gabriele2021} giving predictions for the response using the nonlinear optical kernel formalism. Experimentally, this pump-probe protocol has been used to study LBCO, where both amplification of Josephson Plasma Waves \cite{Rajasekaran2016} and long-lived oscillations in the THz pump-probe signal \cite{Zhang2022} were observed.
\section{Conclusion}
We have explored the nonlinear $c$-axis response of LSCO and have observed THz third-harmonic generation arising from the Josephson plasma mode. The emission from the sample under intense THz radiation displays cubic behaviour indicative of third harmonic generation, which has been shown to be consistent with phase dynamics between the copper-oxygen planes in LSCO. An interesting future direction on the cuprates would be to investigate the relationship between the phase mode and amplitude mode and their potentially coupled contributions to the nonlinear terahertz response.

Acknowledgements: Research at UCSD supported by the U.S. National Science Foundation DMR-1810310 and NASA 80NSSC19K1210. Research  at Columbia is supported by DMR-2210186 and  DMR-2011738. DNB is is the Vannevar Bush Faculty Fellow ONR-VB: N00014-19-1-2630.

\bibliographystyle{apsrev4-1}
\bibliography{library}

\begin{thebibliography}{35}
\bibitem [{Sca()}]{Scale_note}%
  \BibitemOpen
  \href@noop {} {}\bibinfo {note} {The linear spectral amplitude is scaled
  using the reflection coefficient at the minimum field strength. Specifically,
  $E^{*}_{L}(\omega) = E_{L}(\omega) \eta$, where $E_{L}(\omega)$ is the
  spectral amplitude reflected from the LSCO at the lowest incident field.
  $\eta = E^{Au}_{NL}(\omega)/E^{Au}_{L}(\omega)$ where $E^{Au}_{NL}(\omega)$
  and $E^{Au}_{L}(\omega)$ are reference pulses reflected off a gold surface at
  the nonlinear and minimum field strengths, respectively.}\BibitemShut {Stop}%
\end{thebibliography}

\end{document}